\documentclass[conference]{IEEEtran}
\IEEEoverridecommandlockouts
\usepackage[utf8]{inputenc}
\usepackage{cite}
\usepackage{amsmath,amssymb,amsfonts}
\usepackage{algorithmic}
\usepackage{graphicx}
\usepackage{textcomp}
\usepackage{xcolor}
\def\BibTeX{{\rm B\kern-.05em{\sc i\kern-.025em b}\kern-.08em
    T\kern-.1667em\lower.7ex\hbox{E}\kern-.125emX}}
\begin{document}

\title{A Proposal for a Debugging Learning Support Environment for Undergraduate Students Majoring in Computer Science\\
}

\author{\IEEEauthorblockN{Aoi Kanaya}

\IEEEauthorblockA{
\textit{Dept. of Information Technology and Media Design} \\
\textit{Nippon Institute of Technology}\\
Saitama, Japan \\
aoi.kanaya1219@gmail.com
}
\and
\IEEEauthorblockN{Takuma Migo}
\IEEEauthorblockA{
\textit{Dept. of Information Technology and Media Design} \\
\textit{Nippon Institute of Technology}\\
Saitama, Japan \\
migotaku-works0350@ymail.ne.jp
}
\and
\IEEEauthorblockN{Hiroaki Hashiura}
\IEEEauthorblockA{
\textit{Dept. of Data Science} \\
\textit{Nippon Institute of Technology}\\
Saitama, Japan \\
hashiura@nit.ac.jp
}
}

\maketitle

\begin{abstract}

  In software development, encountering bugs is inevitable. However, opportunities to learn more about bug removal are limited. When students perform debugging tasks, they often use print statements because students do not know how to use a debugger or have never used one.

  In this study, among various debugging methods, we focused on debugging using breakpoints. We implemented a function in Scratch, a visual programming language, that allows for self-learning of correct breakpoint placement and systematic debugging procedures.

  In this paper, we discuss experimental results that clarify the changes that occur in subjects when they learn debugging in Scratch.

\end{abstract}

\begin{IEEEkeywords}
Software Engineering, Debugging, Scratch
\end{IEEEkeywords}

\section{Background}

In software development, encountering bugs is inevitable. Therefore, it is desirable to have multiple means of bug fixing, so that, if a bug cannot be fixed with one method, it can be fixed using another. A debugger in an IDE can be considered an effective means of bug fixing. However, content related to debugging is not actively covered in classes, including at our university. In addition, in programming classes, print statements (print debugging) are often used when debugging. From these facts, the need to learn how to use a debugger with breakpoints and systematic debugging procedures\cite{debugging,wolf} has become clear.

\section{Objective}

This research aims to provide learning support for information system students who do not know how to use a debugger by teaching them how to use a debugger with breakpoints and systematic debugging procedures.

\section{Proposed Method and Implementation}

\begin{figure}[htbp]
  \centerline{\includegraphics[width=0.5\linewidth]{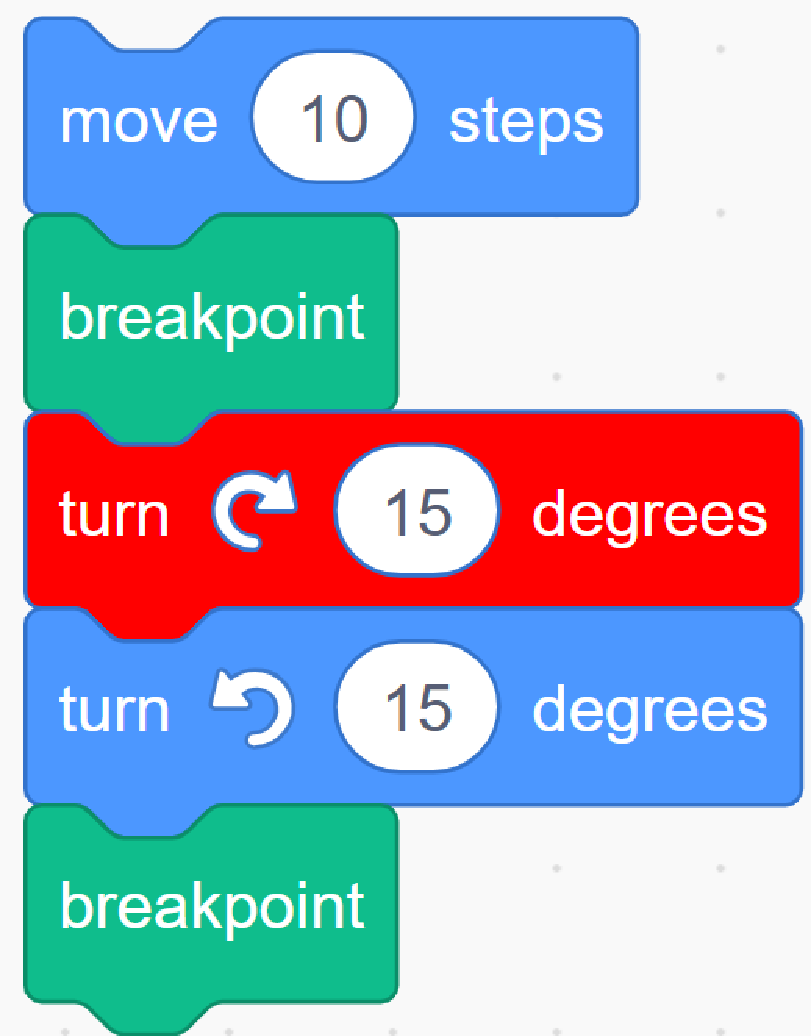}}
  \caption{The feature implemented in Scratch}
  \label{block}
\end{figure}

In this study, as a self-learning support environment for debugging, we added the functionality of breakpoints and step execution to Scratch\cite{Scratch}, as well as the functionality to display paused locations in red. These features aim to bring Scratch closer to the operation of a debugger in a general IDE. This makes it easier to transfer learning from Scratch\cite{Scratch} to other programming environments. Figure \ref{block} shows the implementation of the proposed method.

To verify whether this feature has a positive impact on the subjects, we set the following Research Questions (RQs):

RQ1: Is there a change in the use of debugger functions between tool and non-tool users?

RQ2: Is there a change in debugging procedures between tool users and non-tool users?

RQ3: Can tool users appropriately use the debugger?

\begin{figure}[htbp]
  \centerline{\includegraphics[width=0.7\linewidth]{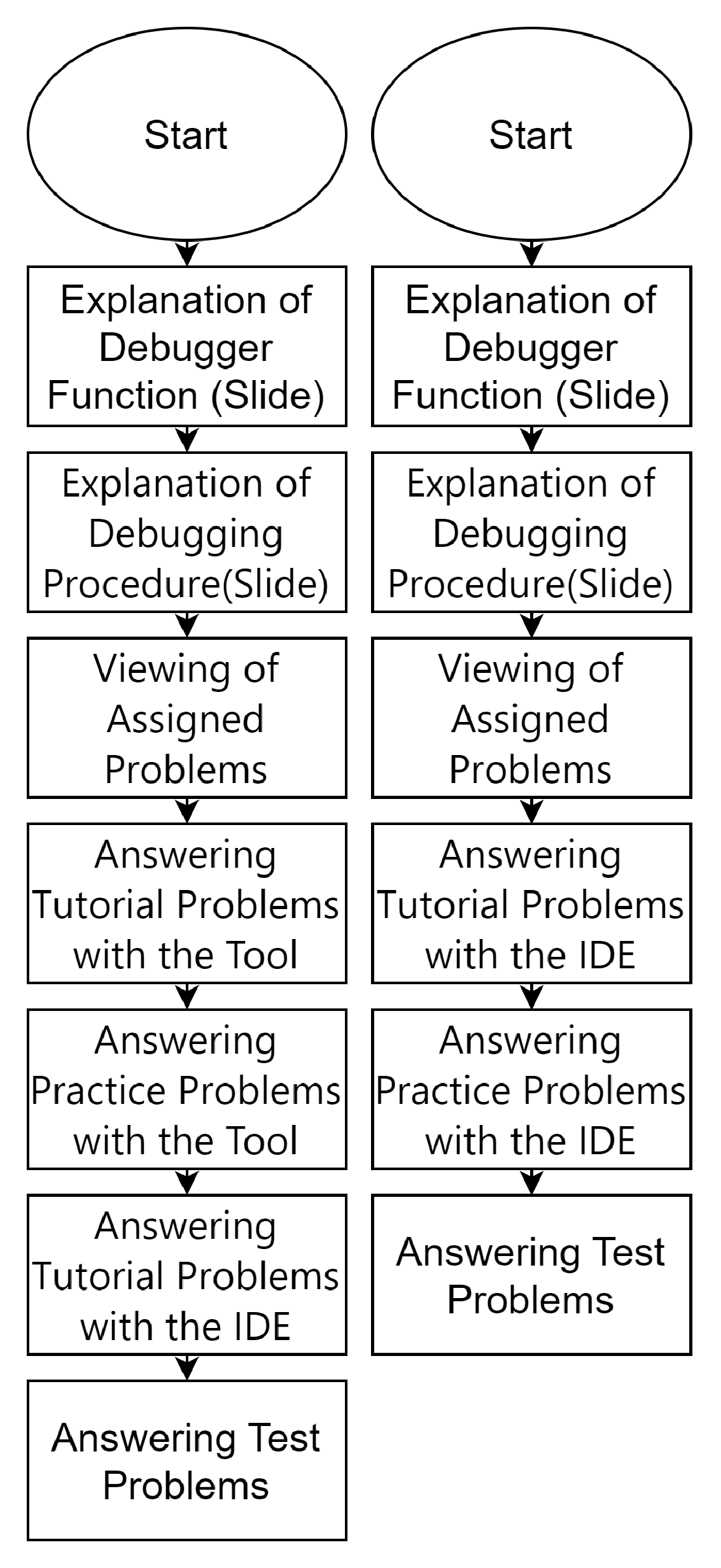}}
  \caption{The process followed by Group A (left) and Group B (right)}
  \label{process}
\end{figure}

An experiment was conducted with 18 undergraduate stu- dents, 1 graduate student, and 1 research student from the Nippon Institute of Technology. The subjects had completed “Programming I” and “Programming II” in their undergraduate studies. The experiment randomly assigned the subjects to a group that learned using the proposed tool (hereafter, Group A) and a group that learned using a general IDE (hereafter, Group B). The ﬂow of the experiment for each group is shown in Figure \ref{process}.

The debugging procedure given to the subjects in the experiment is as follows:

\begin{enumerate}
  \item Confirm the occurrence of the bug
  \item Read the entire program and understand its content
  \item Use breakpoints to intervene in the variables/processes of interest
  \item Predict the behavior of parts where breakpoints are used
  \item Fix the bug when prediction and actual behavior do not match
  \item Repeat steps 3 to 6 until the bug that is occurring is fixed
\end{enumerate}

In this study, debugging refers to the ability of learners to use breakpoints to intervene in the process where the variable, which they predict to be the cause of the bug, is used, following the above procedure. The subjects for evaluating the RQ were the practice and test problems of each group, and the debugger functions and debugging procedures of the subjects were visually confirmed from the video recording. The evaluation items for the debugger function were “breakpoint”, “continue”, “step over”, “step in”, “step out”, and “watch expression”, and if the use was confirmed at least once, it was considered to meet the evaluation items.

The evaluation items for the debugging procedure are steps 3-5. If the following actions are performed: “inserting a breakpoint into the process”, “there is an intention to use the breakpoint”, and “bug fixing”, it is considered that the evaluation items have been met.

\section{Results and Discussion}

Regarding RQ1, a significant difference was observed in the “continue” and “step in” of the practice problems (continue: $\alpha=0.05$, $p=0.0558$) (step in: $\alpha=0.05$, $p=0.0389$). The results where a significant difference was observed are presented in Tables \ref{tab:Count of “Continue” Practice Problems} and Table \ref{tab:Count of Step In Practice Problems Evaluator A}, respectively.

\begin{table}[htbp]
  \centering
  \caption{Count of “Continue” Practice Problems}
  \label{tab:Count of “Continue” Practice Problems}
  \begin{tabular}{cccc}
  \hline
  \multicolumn{4}{c}{Continue}                                            \\ \hline
  \# & Usage                & With tool learning & Without Tool Learning \\ \hline
  1  & Used                  & 0                  & 6                     \\ \hline
  2  & Not Used              & 10                 & 4                     \\ \hline
  3  & p-value (Evaluator A) & 0.055829295        & \multicolumn{1}{l}{}  \\ \hline
  \end{tabular}
\end{table}

\begin{table}[htbp]
    \centering
    \caption{Count of “Step In” Practice Problems (Evaluator A)}
    \label{tab:Count of Step In Practice Problems Evaluator A}
    \begin{tabular}{cccc}
    \hline
    \multicolumn{4}{c}{Step   in}                                           \\ \hline
    \# & Usage                & With tool learning & Without Tool Learning \\ \hline
    1  & Used                  & 10                 & 5                     \\ \hline
    2  & Not Used              & 0                  & 5                     \\ \hline
    3  & p-value (Evaluator A) & 0.038867104        & \multicolumn{1}{l}{}  \\ \hline
    \end{tabular}
\end{table}

For items where a significant difference was recognized, “Continue”, this tool has not been implemented. Therefore, the value of an item with tool learning is 0. In addition, for “Step In”, it is believed that this result was obtained because the operation to execute the step in this tool was not the generally used step over in the debugger but adopted Step In. Regarding RQ2, no items showed a significant difference in both practice problems and tests. RQ3 did not show a significant difference between RQ1 and RQ2, so it can be concluded that groups A and B can use the debugging work equally.

The reason for these results could be that the subjects
were somewhat familiar with traditional IDEs, many of them had little experience with Scratch. Therefore, it is possible that learning without using this tool required no process to become proficient (understand usage and specifications) in Scratch, and they could immediately start debugging work. In addition, this may have caused confusion for subjects who had already somewhat mastered programming languages such as C and Java, where the array index starts from 0, due to issues with Scratch’s specifications, such as the array index starting from 1. The problems were designed to be the same for groups A and B, but group B grasped the problems faster. This could be because the program written in Scratch appeared to have more content due to the constraint of being a diagrammatic language than the program written in the IDE.

In this experiment, the recordings were tallied by visual inspection; thus, there may have been a problem with the reproducibility of the tallying method. To clarify this, tallying was performed by two people, and the differences were compared. As a result, it was found that there were places where a difference of 1-2 in the numbers occurred depending on the tally person. Most of these differences did not affect the evaluation results, but it became clear that they affected the “Step In” of the practice problems. The results are presented in Table \ref{tab:Count of Step In Practice Problems Evaluator B}.

\begin{table}[htbp]
  \centering
  \caption{Count of “Step In” Practice Problems (Evaluator B)}
  \label{tab:Count of Step In Practice Problems Evaluator B}
  \begin{tabular}{cccc}
  \hline
  \multicolumn{4}{c}{Step   in}                                           \\ \hline
  \# & Usage                & With tool learning & Without Tool Learning \\ \hline
  1  & Used                  & 10                 & 5                     \\ \hline
  2  & Not Used              & 0                  & 5                     \\ \hline
  3  & p-value (Evaluator A) & 0.038867104        & \multicolumn{1}{l}{}  \\ \hline
  \end{tabular}
\end{table}

In this experiment, the presence or absence of a significant difference was limited to “Step In”, but the fact that a difference in the count number is a problem; thus, it is necessary to revise the evaluation criteria in future tasks. Also, it became clear that the reason for such a difference was due to the visual check of the recording; thus, it was necessary to keep the usage as a log.

\section{Related Work}

In related work, the learning system proposed by Yamamoto et al.\cite{batari_debug} supports the learning of debugging procedures. When learners solve problems, the current debugging procedure is highlighted. When the problems set for each procedure are answered, they proceed to the next procedure and continue learning using the tool. However, there has been no discussion about learning debugging procedures and debuggers in visual programming languages.

\section{Conclusion}

From the experimental results, it was found that there was no change in the learning of how to use the debugger function with breakpoints and systematic debugging procedures, regardless of whether the tool used in this study was used or not. As a future task, it is suggested to investigate the reason why no useful change occurred in learning with this tool.

\section*{Acknowledgment}

Part of this research was supported by JSPS KAKENHI Grant Number 21K12179.

\bibliographystyle{jIEEEtran}  
\bibliography{IEEEfull}  

\end{document}